\documentclass{moriond}

\usepackage{lineno}

\def\PLB{{\em Phys. Lett.}  B}
\def\PRL{\em Phys. Rev. Lett.}

\def\be{\begin{equation}}
\def\ee{\end{equation}}
\def\bea{\begin{eqnarray}}
\def\eea{\end{eqnarray}}

\newcommand{\Photo}{}

\newcommand{\epem}{\rm e^+e^-}
\newcommand{\ep}{\rm e^-{\rm p}}

\newcommand{\plead}{p--Pb}
\newcommand{\leadlead}{Pb--Pb}

\newcommand{\pt}{p_{\rm T}}

\newcommand{\ssqrt}{\sqrt{s}}

\newcommand{\m}{{ m}}

\newcommand{\rplead}{R_{\rm pPb}}

\newcommand{\ccbar}{\rm c\bar{c}}

\newcommand{\sNN}{s_{\rm NN}}

\newcommand{\Lc}{\Lambda_{\rm c}^{+}}

\newcommand{\Dzero}{{\rm D}^0}

\newcommand{\Ds}{{\rm D_s^+}}

\newcommand{\Xic}{\Xi_{\rm c}}
\newcommand{\Xiczero}{\Xic^0}

\begin{document}
\vspace*{4cm}
\title{Studies on the hadronization of charm and beauty quarks}

\author{Mattia Faggin, on behalf of the ALICE Collaboration}

\address{Department of Physics and Astronomy ``Galileo Galilei''\\
University of Padova, Italy}

\maketitle\abstracts{
In this contribution, the latest results on hadronization studies of charm and beauty quarks obtained with the data collected with the ALICE experiment at the LHC are presented. Measurements of prompt and non-prompt charm-hadron production in pp, \plead\ and \leadlead\ are shown. The results are also compared with theoretical models that consider different implementations of the heavy-quark hadronization across collision systems.}

\section{Heavy quarks: a unique probe of high-density QCD}
Given their large mass ($\m_{\rm c}\approx 1.3$ GeV/$c^2$, $\m_{\rm b}\approx 4.2$ GeV/$c^2$), 
charm and beauty quarks at the LHC are produced mainly in the hard scattering processes among partons of the colliding protons or nuclei. In addition, in \leadlead\ collisions charm and beauty quarks are produced before the onset of the quark-gluon plasma (QGP), a deconfined state of the strongly-interacting matter characterized by partonic degrees of freedom. Charm and beauty quarks are unique probes of the QGP, since they experience its full evolution, and of hadronization mechanisms in any hadronic collision system. Since the flavour is retained through the collision evolution, the charm and beauty-quark hadronization can be tested experimentally via the measurement of charm and beauty-hadron production. Such measurements in \leadlead\ collisions aim to study the properties of the QGP and the effects on the heavy-quark dynamics and hadronization. Heavy-flavour hadron production in \plead\ collisions is of great interest to study the influence of the so-called cold nuclear matter effects, induced by the presence of nuclei in the initial state. 
An important reference to the studies in \plead\ and \leadlead\ is provided by pp collisions. In addition, measurements in such collision system are very important per se, because they provide fundamental tests of perturbative QCD (pQCD) calculations.

Calculations of heavy-flavour hadron production in $\epem\to Q\bar{Q}$ collisions are usually based on a factorization approach, with the production cross section of charm and beauty hadrons described as the product of:~(i) the cross section of the hard scattering process between the two leptons, calculated in pQCD; (ii) the fragmentation functions (FF), which encode the information on the hadronization and quantify the probability for a heavy-flavour quark to produce a hadron carrying a given fraction of the original quark momentum. 
This approach is adopted also in pp collisions with the addition of the parton distribution functions, which quantify the probability to pick-up a parton of a given momentum from the colliding protons. Model calculations based on a factorization approach, and relying on FF that are constrained from $\epem$ and $\ep$ measurements, well describe the charm and beauty meson production at the LHC \cite{alice_pnpDpp5TeV}. 
Heavy-flavour mesons are also measured in \leadlead\ collisions, where their production is significantly influenced by QGP-induced effects, as strangeness enhancement and coalescence of heavy quarks with light quarks from the bulk \cite{HeRapp_strangeness_coalescence}. The interplay between these two phenomena explains the larger nuclear modification factor of $\Ds$ mesons in $2<\pt<8$~GeV/$c$ compared to that of non-strange D mesons \cite{alice_DsPbPb5TeV}.
%
%

While measurements of heavy-flavour meson production are well described by model predictions, models based on a factorization approach do not describe the baryon production in pp collisions at the LHC, where a significant enhancement of the baryon-to-meson ratios compared to those in $\epem$ collisions is observed \cite{alice_LcScpp13TeV,alice_Xicpp13TeV}. Moreover, a different trend as a function of the event multiplicity is observed for $\pt$-differential ratios including mesons and baryons \cite{alice_DLcvsMultpp13TeV}. Therefore, the charm and beauty hadronization mechanisms from $\epem$ to \leadlead\ collisions are still a puzzle. To solve it, more precise and differential measurements are needed.

\section{The ALICE experiment and recent experimental results}
A Large Ion Collider Experiment (ALICE) is a general-purpose experiment conceived to study the properties of the strongly-interacting matter. It comprises a central barrel covering the pseudorapidity interval $|\eta| < 0.9$, where charged particles are tracked and identified via the Inner Tracking System (ITS), the time Projection Chamber (TPC) and the Time-Of-Flight (TOF) detector. Profiting of the excellent pointing resolution to the primary vertex of some tens of microns provided by the two innermost layers of ITS, the decay point of heavy-flavour hadrons (secondary vertex) could be resolved from the beam interaction point (primary vertex).

\begin{figure}[t]
	\centering
	\includegraphics[width=0.45 \textwidth]{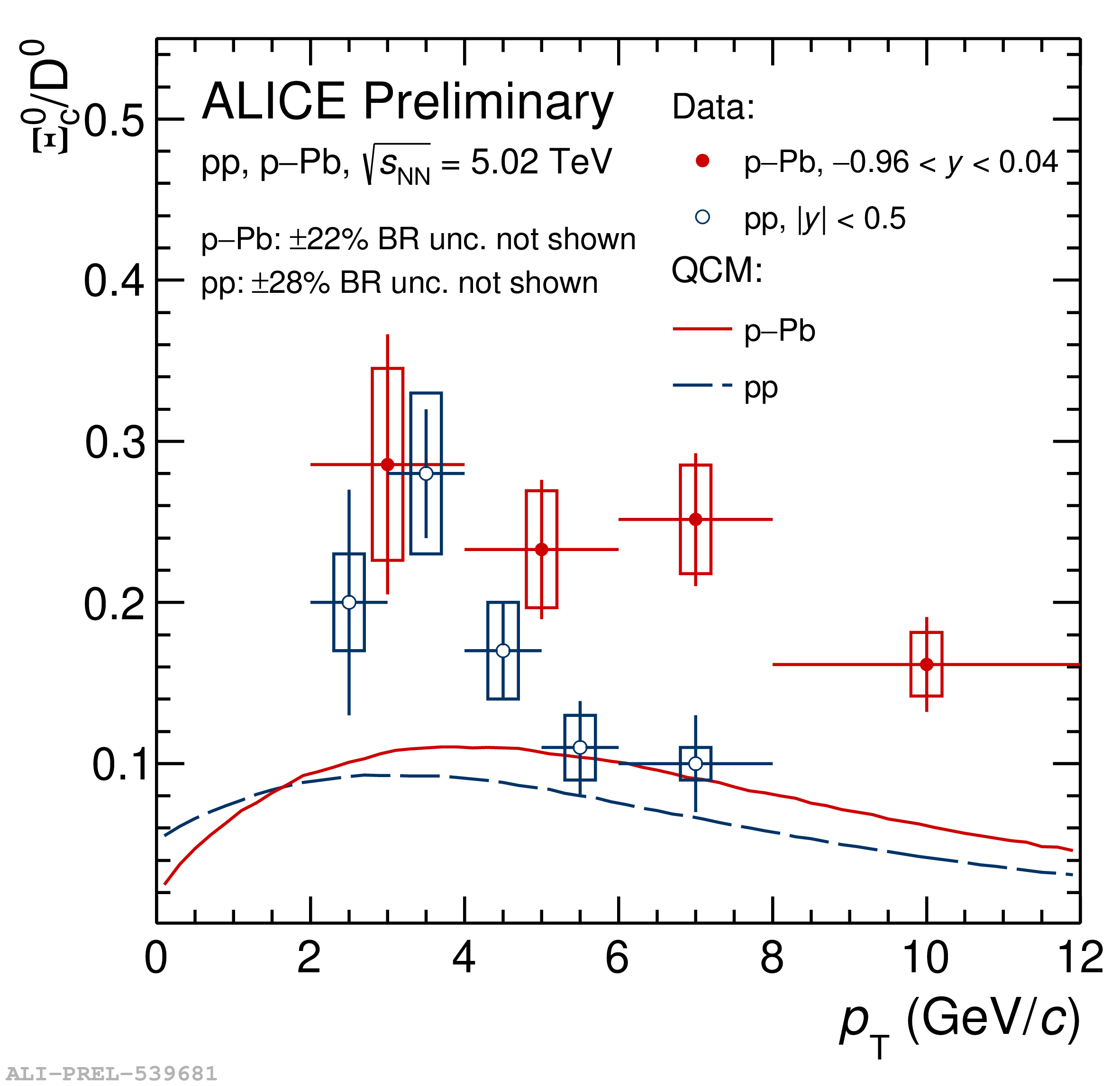} \quad
	\includegraphics[width=0.45 \textwidth]{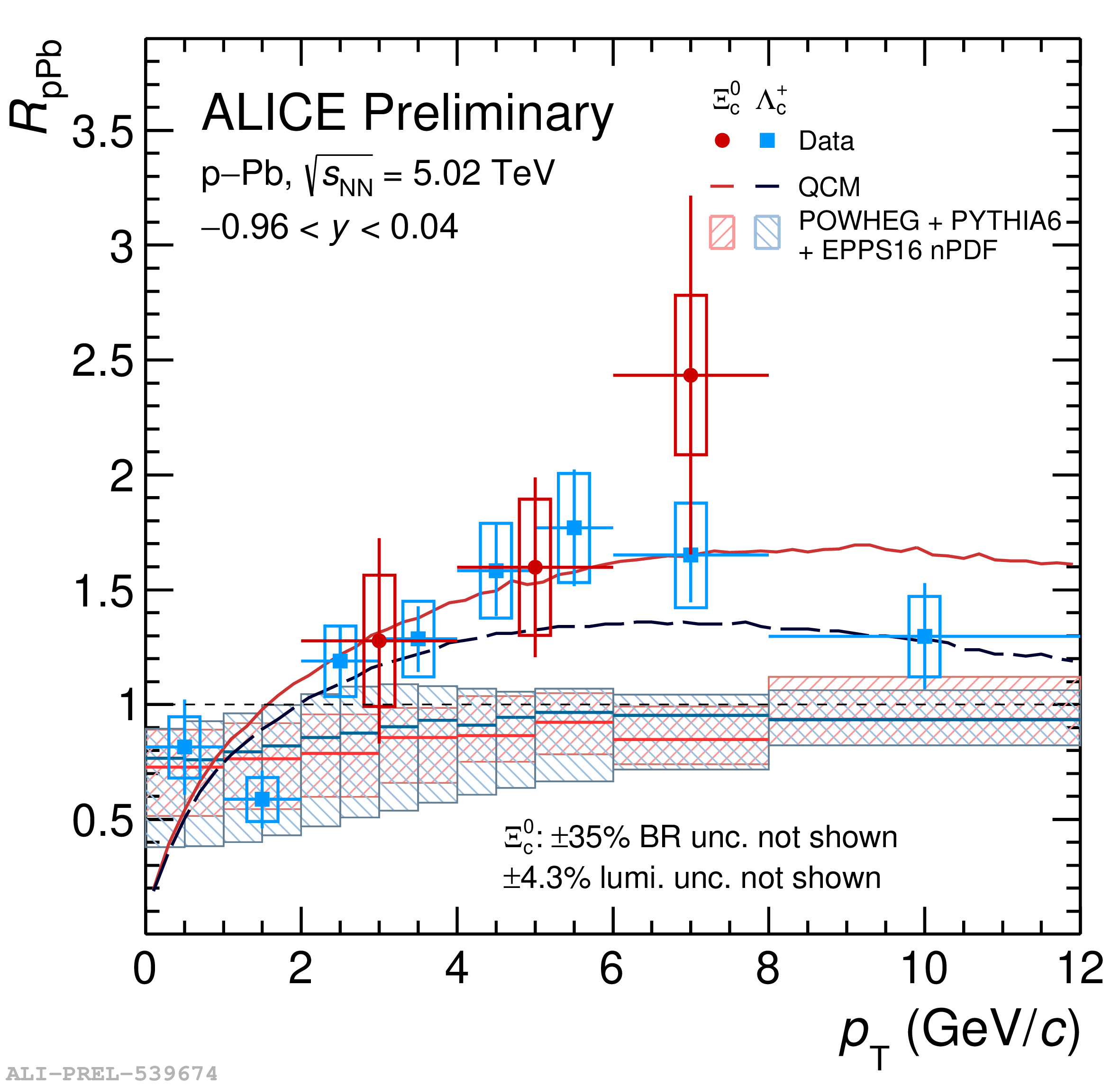} \\
	\caption{Left: $\pt$-differential $\Xiczero/\Dzero$ ratio at midrapidity in pp and \plead\ collisions at $\sqrt{\sNN}=5.02$ TeV. Right: nuclear modification factor ($\rplead$) of $\Lc$ and $\Xiczero$ baryons at midrapidity in \plead\ collisions.}
	\label{fig:Xic}
\end{figure}

\begin{figure}[t]
	\centering
	\includegraphics[width=0.44 \textwidth]{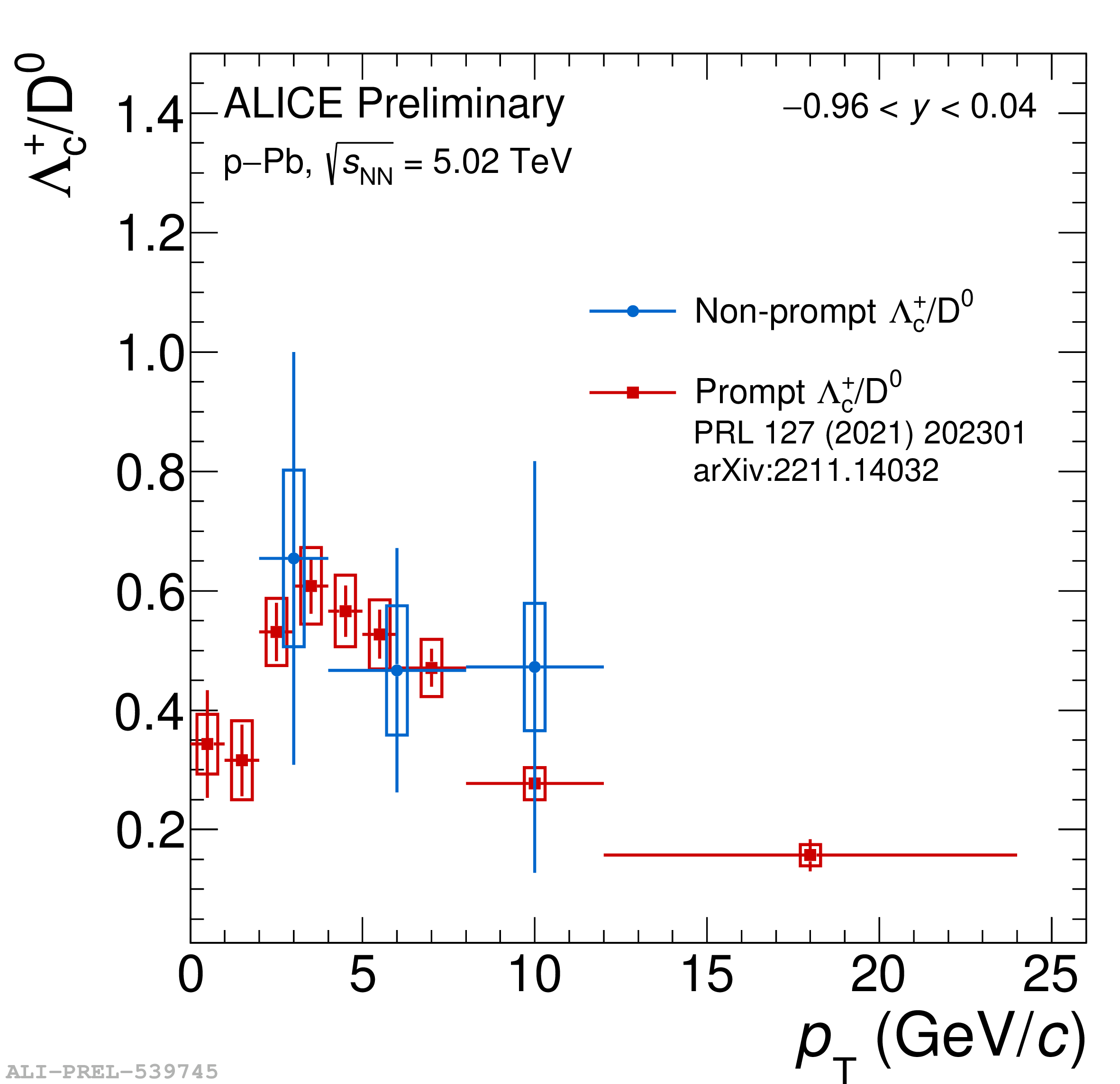} \quad
	\includegraphics[width=0.50 \textwidth]{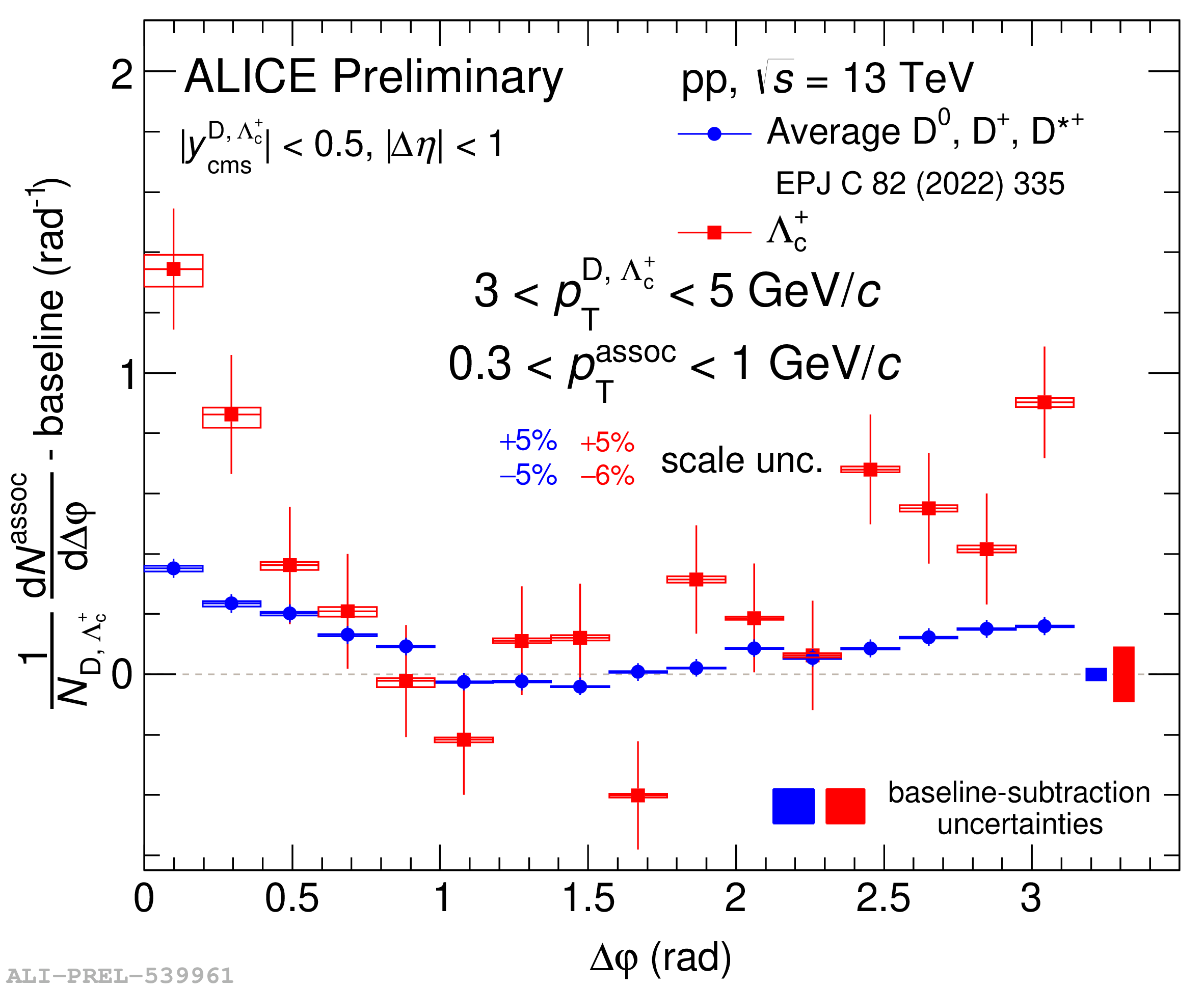} \\
	\caption{Left: prompt and non-prompt $\pt$-differential $\Lc/\Dzero$ ratio at midrapidity in \plead\ collisions at $\sqrt{\sNN}~=~5.02$~TeV. Right: $\Lc$-hadron and $\Dzero$-hadron azimuthal correlation functions at midrapidity in pp collisions at $\ssqrt=13$ TeV.}
	\label{fig:Lc}
\end{figure}

ALICE measured the $\pt$-differential production of prompt and non-prompt D mesons in pp collisions at $\sqrt{s}=5.02$ TeV \cite{alice_pnpDpp5TeV} and 13 TeV for $|y|<0.5$. The latter measurements allow an experimental access to the beauty-meson production mechanisms. The meson-to-meson ratios do not show any significant $\pt$-dependence, being compatible with those measured in $\epem$ collisions, and they are correctly described by pQCD-based models \cite{FONLL} employing a factorization approach and FF constrained from $\epem$ and $\ep$ measurements \cite{alice_pnpDpp5TeV}. On the other hand, the $\Lc/\Dzero$ ratio in pp collisions for $|y|<0.5$ shows an enhancement by a factor of about 5 at low $\pt$ compared to the result in $\epem$ collisions, and a decreasing trend as a function of $\pt$. Several model calculations \cite{pythiaCR,Catania,HeRapp,QCMpp}, implementing hadronization mechanisms different from the Lund fragmentation model, agree with the measurement. Also in \plead\ collisions the $\Lc/\Dzero$ ratio is larger than the measurement in $\epem$ collisions. In addition, the average $\pt$ of the $\Lc$-baryon is observed to be larger in \plead\ than in pp collisions, suggesting a different $\pt$ redistribution for baryons and mesons from pp to \plead\ collisions \cite{alice_Lcpt0pppPb5TeV}. 
ALICE performed also the first measurements at midrapidity of the $\Xiczero/\Dzero$ ratio in pp \cite{alice_Xicpp5TeV} and \plead\ collisions at the LHC (Fig.~\ref{fig:Xic} left). They are both larger than the results in $\epem$, with none of the models mentioned above being able to describe the ratio in pp collisions. The $\Xiczero/\Dzero$ ratio in \plead\ collisions is higher than the result in pp collisions for $\pt>6$ GeV/$c$, and the quark-combination mechanism (QCM) model \cite{QCMXicpPb} underestimates the data over almost the full $\pt$ range (Fig.~\ref{fig:Xic} left). In addition, the $\rplead$ of $\Lc$ and $\Xiczero$ show the same $\pt$ dependence and magnitude within uncertainties, being larger than unity for $4<\pt<8$~GeV/$c$ (Fig.~\ref{fig:Xic} right). This latter result supports the hypothesis of a $\pt$ redistribution in \plead\ collisions common to all charm baryons compared to pp collisions.
The baryon enhancement observed from the $\pt$-differential charm baryon-to-meson ratio significantly influences also the charm fragmentation fractions at midrapidity, measured for the first time in both pp\cite{alice_charmFFppTeV} and \plead\ collisions. The $\Lc$ fragmentation fraction in pp collisions at the LHC is larger by a factor of about 3 compared to the results in $\epem$ and $\ep$ collisions, and a reduction of about 1.4-1.6 for the D mesons is observed. The fragmentation fractions, as well as the $\ccbar$ production cross section $\sigma(\ccbar)/A$, with $A$ the mass number of the projectile, measured for the first time at midrapidity ($|y|<0.5$) in both collision systems as the sum of the ground-state hadron cross sections, do not show any significant system dependence. 
An enhancement of the baryon-to-meson yield ratio with respect to $\epem$ collisions is observed also for the beauty sector, accessed experimentally via the measurement of non-prompt charm hadrons. ALICE measured for the first time the production of non-prompt $\Dzero$ and $\Lc$ hadrons and their ratio at midrapidity down to $\pt=2$~GeV/$c$ in pp and \plead\ collisions (Fig.~\ref{fig:Lc} left). In both collision systems, the prompt and non-prompt $\Lc/\Dzero$ ratios show a similar enhancement compared to $\epem$ results, and within the current uncertainties no system dependence is observed for the non-prompt $\Lc/\Dzero$ ratio.
To shed light on the possible presence of similar effects among pp, \plead\ and \leadlead\ collisions, the prompt $\Ds/\Dzero$ and $\Lc/\Dzero$ ratios were also studied as a function of event multiplicity \cite{alice_DLcvsMultpp13TeV}. While the $\Ds/\Dzero$ $\pt$-differential ratio does not show any significant $\pt$ and multiplicity dependence, the $\Lc/\Dzero$ ratio decreases with $\pt$ and shows an increase in $1<\pt<24$ GeV/$c$ with a significance of $5.3\sigma$. This behaviour is qualitatively described by PYTHIA predictions only if modes employing colour reconnection mechanisms beyond leading colour approximation \cite{pythiaCR} are considered, while the default Monash tune underestimates the data. The prompt $\Lc/\Dzero$ ratio further increases at intermediate $\pt$ in \leadlead\ collisions \cite{alice_LcPbPb5TeV}, though the $\pt$-integrated ratio does not show any significant dependence on event multiplicity, and it is compatible among the different collision systems. This result further supports scenarios explaining the increase at intermediate $\pt$ as a redistribution of $\pt$ acting differently for mesons and baryons in pp, \plead\ and \leadlead\ collisions. This behaviour is not explained by model calculations, which are not able to describe either the magnitude or the multiplicity dependence of the measurement.
Further insights on the charm-quark hadronization at the LHC are provided by the first measurement of azimuthal-correlation function between $\Lc$ baryons and charged particles at midrapidity in pp collisions at $\ssqrt=13$ TeV (Fig.~\ref{fig:Lc} right). The near-side and away-side peak yields for $3<\pt(\Lc)<5$ GeV/$c$ and associated particles within $0.3<\pt<1$ GeV/$c$ are larger than those of the D-meson correlation function in the same collision system \cite{alice_Dhpp13TeV}, a feature not described by PYTHIA simulations. Possible explanation of this difference could be related to the impact of decays of higher-mass charm baryons, as prescribed in Ref. \cite{HeRapp}, or to the softer $\Lc$ fragmentation, as pointed-out by the measurement of $\Lc$-tagged jets in the same collision system \cite{alice_Lcjetspp13TeV}. 
Finally, ALICE performed the first measurement of $\rm D_{s1}^+(2536)$ and $\rm D_{s2}^{*+}(2573)$ resonances in pp collisions, which provide further pieces of information with respect to those of ground-state charm mesons to investigate the charm-quark hadronization and its dependence on collision system and multiplicity. 
While no clear multiplicity dependence is observed for the $\rm D_{s1}^+(2536)/\Ds$ ratio, the $\rm D_{s2}^{*+}(2573)/\Ds$ result shows a hint of decreasing trend, though with large uncertainties. If confirmed by more precise measurements, such result might shed light on the interplay between the different hadron lifetime and the hadronic rescattering, which is expected to influence their abundance.
 
\vspace{-\baselineskip}
In summary, to solve the puzzle on the heavy-quark hadronization at the LHC, more precise measurements of charm and beauty hadron production are required. During the Run 3, ALICE will collect a factor of hundreds more data than in Run 1 and Run 2, with an improved track impact parameter to the primary vertex thanks to the upgraded ITS. These data will pave the way to precise measurements of more differential observables (heavy-flavour jets, correlations, \dots) and to perform beauty measurements via the direct reconstruction of beauty hadrons.

\end{document}